\documentstyle[aps,12pt,manuscript]{revtex}
\begin{document}
\preprint{INJE--TP--98--1, hep--th/9801152}

\title{Stability analysis and absorption cross--section in 
three--dimensional black string}

\author{ H.W. Lee$^1$, N.J. Kim$^1$, Y. S. Myung$^1$ and Jin Young Kim$^2$}
\address{$^1$Department of Physics, Inje University, Kimhae 621-749, Korea\\ 
$^2$Division of Basic Science, Dongseo University, Pusan 616-010, Korea}

\maketitle

\begin{abstract}
It is shown that all string fields except dilaton are 
non--propagating in the $(2+1)$--dimensional
black string.  One finds that the perturbation around the black string 
reveals a mixing between the dilaton and other fields.  Under the 
new gauge(dilaton gauge), we disentangle this mixing and obtain 
one decoupled dilaton equation.  It turns out that this black string 
is stable.  
From the scattering of dilaton off the neutral black string($N=0$), 
we find the absorption cross--section.
Further the absorption cross--section for minimally coupled 
scalar is obtained and we compared it with that of dilaton.
\end{abstract}
\vfill
Compiled at \today.

\section{Introduction}
Black strings play an important role in understanding 
the microscopic
 origin for the black hole entropy. The statistical interpretation 
of the Bekenstein-Hawking entropy 
was made first in the  extremal or near-extremal five-dimensional 
(5D) black holes\cite{Str96}. 
The microstates of 5D extremal black hole 
arise mainly from the fields moving around a circle in the 
internal dimension. In order to
understand this situation, it is useful to take this internal 
direction as a spacetime direction 
explicitly. This is the six-dimensional (6D) black string\cite{Hor96}. 
The entropy calculation of 6D black string gives us the same result 
as in the 5D black hole.

On the other hand, to calculate the entropy of 
Schwarzschild black hole, we need
a nonperturbative formulation of string theory, M--theory.
This is because the Schwarzschild black hole belongs to the non--extremal one.
Recently, there have been many attempts along this 
direction\cite{Ban97}. The idea is to map 
the Schwarzschild black hole to extremal
 or near-extremal configurations, where the entropy counting might 
be possible\cite{Das97}.
It is shown that black strings give rise to
 Schwarzschild black hole with the compactification radius 
$(R_1)$ or charged dilaton black hole
with the radius ($R_2$), when compactified to lower dimensions
 before or after boosting($\alpha$) along uncompactified direction. 
One finds $R_1=R_2 \cosh \alpha$ and the Newton constant 
$G_1^N = G_2^N/\cosh \alpha$. Thus the entropy of two holes 
are the same. 
Further a 7D Schwarzschild black hole is obtained as a compactification of a 
 three brane in 11D supergravity\cite{Eng98}. And we relate it to a 
charged black hole with the same statistical
 entropy. The charged black hole will be found from 
subjecting the three brane to a boost 
in uncompactified space--time, followed by Kaluza-Klein compactification. 
A near--extremal charged hole is defined such that 
the Schwarzschild radius remains arbitrarily large at 
infinite boost($\alpha \rightarrow \infty$). This case is
used to derive the entropy of neutral Schwarzschild hole 
from the charged ones.

In this paper, we study the propagation of string fields in 
the three--dimensional(3D) black string background 
with the new(dilaton) gauge. 
 The 3D black string is used as a toy model for investigating 
higher--dimensional black strings\cite{Gib95}. 
Especially, the s--mode approximation of higer-dimensional 
black string is exactly the same 
as in the 3D black string\cite{Gre94}. Here we are interested in the 
stability of 3D balck string. 
The transition between Schwarzschild black holes to black branes in 
M--theory corresponds to the balck hole--black string transition in a 
boosted frame\cite{Ban97,Das97}. The latter is related to the 
instability of a black string\cite{Gre94}. In this sense, it is very 
important to analyze the stability of a black string.
Further we investigate the 
dynamical behavior(absorption cross--section=greybody factor) of 
3D black string rather than the static 
behavior(entropy)\cite{Mal97,Cal97}.  Apart from counting the microstates 
of black holes, the dynamical behavior is also an important issue. This is 
so because the greybody factor for the black hole arises as a consequence 
of the gravitational potential barrier surrounding the horizon.  
That is, this is an effect of space--time curvature. Together with the 
Bekenstein--Hawking entropy, this seems to be the strong hint of a deep 
and mysterious connection between curvature and statistical mechanics.

The organization of this paper is as follows. 
In Sec.II we set up the equations of motion for the
3D black string and linearize these equations around the background solution.
 And we introduce both the 
transverse gauge and the new (dilaton) gauge.
In Sec.III, we study the propagation of string fields with 
the gauge conditions. It turns out that
 a physically propagating mode is just the dilaton. 
Further we show that the 3D black string is stable
against the external perturbation. Section IV is concerned with 
the dynamical behavior of the neutral
(N=0) black string. We derive the absorption cross section 
from the scattering process of dilaton off
 the 3D black string. Finally we discuss our results in Sec.V.

\section{Formalism}
Let us start with the $\sigma$-model action of  string theory\cite{Cal85} 

\begin{equation}
S_\sigma = {-1 \over 4 \pi \alpha^\prime} \int d^2 \xi 
\left (\sqrt{\gamma} 
 \gamma^{ab} \partial_a X^\mu \partial_b X^\nu g_{\mu\nu} 
     +  \epsilon^{ab} \partial_a X^\mu \partial_b X^\nu B_{\mu\nu} 
     - {1 \over 2} \alpha^\prime \sqrt{\gamma}R^{(2)} \Phi  \right ),
\label{string_action}
\end{equation}
where $R^{(2)}$ is the Ricci curvature of the world sheet, 
$B_{\mu\nu}$ two--form field, and $\Phi$ dilaton.
The conformal invariance requires the $\beta$-function equations 

\begin{eqnarray}
R_{\mu\nu} - \nabla_\mu \nabla_\nu \Phi 
-{1\over 4} H_{\mu \rho \sigma} H_\nu^{\rho \sigma} &=& 0,
\label{eq_graviton} \\
\nabla^2 \Phi + (\nabla \Phi)^2 - {8 \over k} - {1 \over 6} H^2   &=& 0,
\label{eq_scalar1}  \\
 \nabla_\mu H^{\mu \nu \rho} + (\nabla_\mu \Phi) H^{\mu \nu \rho} &=& 0,
\label{eq_anti}
\end{eqnarray}
where $ H_{\mu \nu \rho} = 3 \partial_{[\mu} B_{\nu \rho]}$ is the 
Kalb--Ramond field corresponding to
$ B_{\nu \rho}$.
The above equations are also derived from the requirement that the 
fields must be an extremum
of the low-energy string action in string frame,

\begin{equation}
S_{l-e} = \int d^3 x \sqrt{-g} e^{\Phi}
   \big \{ R + (\nabla \Phi)^2 + {8 \over k} - {1 \over 12} H^2  \big \}.
\label{action}
\end{equation}

The static black string solution to 
(\ref{eq_graviton})--(\ref{eq_anti}) is found to be

\begin{eqnarray}
&& \bar H_{rtx} = {Q \over r^2}, ~~~~~~\bar 
\Phi = \ln r - {1 \over 2} \ln ({k\over 2}),
      \nonumber   \\
&& \bar g_{\mu\nu} =
 \left(  \begin{array}{ccc} - (1 - {M \over r}) & 0 & 0  \\
                             0 & (1 - {N \over r}) & 0  \\
    0 & 0 & {k \over 8 r^2} (1 - {M \over r})^{-1} 
(1 - {N \over r})^{-1} \end{array}   \right),
\label{bck_metric}
\end{eqnarray}
with $N \equiv Q^2/M (M>N)$.  
The Christoffel symbols are given by

\begin{eqnarray}
&&\Gamma^t_{tr} = {1 \over 2} \Big( { 1 \over r - M} - {1 \over r} \Big),
    \nonumber   \\
&&\Gamma^x_{xr} = {1 \over 2} \Big( { 1 \over r - N} - {1 \over r} \Big),
    \nonumber   \\
&&\Gamma^r_{tt} = {4 \over k} M \Big( 1 - { M \over r } \Big)  
                            \Big( 1 - { N \over r } \Big),  \nonumber   \\
&&\Gamma^r_{xx} = - {4 \over k} N \Big( 1 - { M \over r } \Big)  
                            \Big( 1 - { N \over r } \Big),  \nonumber   \\
&&\Gamma^r_{rr} = - {1 \over 2} \Big( { 1 \over r - M} + 
                {1 \over r - N} \Big).    
\label{gamma_comp}
\end{eqnarray}
A simple extension of Witten's construction for 
a gauged WZW model also yields the 3D charged black string 
\cite{Hor92}.  
The solution (\ref{bck_metric}) is characterized by three 
parameters: $M$ (mass), 
$Q$ (axion charge per
unit length), and $k$ (cosmological constant). 
For $0 < |Q| < M$, black string is similar to the 
4D Reissner-Nordstr\"om solution. 
In addition to the event (outer) 
horizon ($r_{EH} = M$), there exist
an inner horizon ($r_{IH} = N$).  When $|Q| = M$, 
this becomes the extremal black string.  
Finally, when $|Q| > M$, the spacetime 
has neither a horizon nor a curvature singularity and this case 
is not relevant to us. 

To study the propagation of fields specifically, we introduce 
the small perturbation fields around the background solution 
as \cite{Lee95,Lee97}
\begin{eqnarray}
&&H_{rtx} = \bar H_{rtx} + {\cal H}_{rtx},      \nonumber   \\   
&&\Phi = \bar \Phi + \phi,                      \nonumber   \\  
&&g_{\mu\nu} = \bar g_{\mu\nu} + h_{\mu\nu}. 
\label{perturbation}
\end{eqnarray}
In order to obtain the  equations governing the perturbations, 
we introduce the notation
\begin{equation}
{\hat h}_{\mu\nu} = h_{\mu\nu}-{1 \over 2} {\bar g}_{\mu\nu} h.
\label{h_bar}
\end{equation}
And then one needs to linearize (\ref{eq_graviton})--(\ref{eq_anti}) 
to obtain
\begin{eqnarray}
  \delta R_{\mu\nu} (h) 
- \bar \nabla_\mu \bar \nabla_\nu \phi   
+ \delta \Gamma^\rho_{\mu\nu} (h) \bar \nabla_\rho \bar \Phi
- {1 \over 2} \bar H_{\mu \rho \sigma} {\cal H}_\nu^{ \rho \sigma}   
+ {1 \over 2} \bar H_{\mu \rho \sigma} \bar H_{\nu\alpha}^{~~\sigma} 
h^{\rho \alpha} &=& 0,
\label{lin_graviton} \\
 h^{\mu\nu} \bar \nabla_\mu \bar \nabla_\nu \bar \Phi
+ h^{\mu\nu} \partial_\mu \bar \Phi \partial_\nu \bar \Phi   
+ \bar \nabla_\mu {\hat h}^{\mu\rho}\partial_\rho \bar \Phi
- \bar \nabla^2 \phi 
- 2 \bar g^{\mu\nu} \partial_\mu \bar \Phi \partial_\nu \phi && \nonumber   \\  
~~~~~~~~~~~~~~+ {1 \over 6} \Big \{ 2 \bar H_{\mu \rho \sigma} 
                      {\cal H}^{\mu \rho \sigma}                
     - 3 \bar H_{\mu \rho \sigma}   \bar H^{\alpha \rho \sigma} h^\mu_\alpha 
                                    \Big \} &=& 0, 
\label{lin_scalar} \\
  ( \bar \nabla_\mu + \partial_\mu \bar \Phi ) 
     {\cal H}^{\mu \nu \rho} 
- ( \bar \nabla_\mu h_\beta^{~\nu}) {\bar H}^{\mu\beta\rho}
+ (\bar \nabla_\mu h_\beta^{~\rho}) {\bar H}^{\mu\beta\nu}
- (\bar \nabla_\mu {\hat h}_{~\alpha}^{\mu}) {\bar H}^{\alpha\nu\rho}
      + (\partial_\mu \phi) \bar H^{\mu \nu \rho}
 &=& 0, 
\label{lin_anti}
\end{eqnarray}
where 
\begin{eqnarray}
&&\delta R_{\mu\nu} (h) = 
 - {1 \over 2} \bar \nabla^2 h_{\mu\nu}   
       - {1 \over 2} \bar \nabla_\mu \bar \nabla_\nu h^\rho_{~\rho}
 + {1 \over 2} \bar \nabla^\rho \bar \nabla_\nu h_{\rho\mu}   
 + {1 \over 2} \bar \nabla^\rho \bar \nabla_\mu h_{\nu\rho},   
\label{delR} \\   
&&\delta \Gamma^\rho_{\mu\nu} (h) = {1 \over 2} \bar g^{\rho\sigma} 
( \bar \nabla_\nu h_{\mu\sigma} + \bar \nabla_\mu h_{\nu\sigma} - 
            \bar \nabla_\sigma h_{\mu\nu} ).
\label{delG}
\end{eqnarray}
Here $\delta R_{\mu\nu}$ can be transformed to the Lichnerowicz 
operator\cite{Gre94}
\begin{equation}
\delta R_{\mu\nu} = -{1 \over 2} \bar \nabla^2 h_{\mu\nu} 
 +{\bar R}_{\sigma ( \nu} h^\sigma_{~\mu )}
 -{\bar R}_{\sigma \mu\rho\nu} h^{\sigma\rho}
 + \bar \nabla_{( \nu} \bar \nabla_{|\rho|} {\hat h}^\rho_{~\mu)}.
\label{delR_Lic}
\end{equation}
These are the bare perturbation equations.  We have to examine whether 
there exist any choice of gauge which can simplify 
(\ref{lin_graviton})--(\ref{lin_anti}).  Conventionally, we choose the 
harmonic(transverse) gauge($\bar \nabla_\mu {\hat h}^{\mu\rho} = 0$) to
describe the propagation of gravitons.  Further one requires the traceless 
condition($h=0$), which corresponds to de Donder gauge.  However this does 
not eliminate all of the gauge degrees of freedom for the metric.  There 
still exist the residual gauge transformations,
\begin{equation}
\delta_\xi h_{\mu\nu} = \bar \nabla_\mu \xi_\nu + \bar \nabla_\nu \xi_\mu
\label{residual}
\end{equation}
with $\bar \nabla_\mu \xi^\mu = 0 $ and $\bar\nabla^2 \xi^\mu=0$.  The 
counting of gravitational degrees of freedom is as follows.  A symmetric 
traceless tensor has $D(D+1)/2-1$ in $D$--dimensions.  $D$ of them 
are eliminated by the harmonic gauge condition.  Also $D-1$ are 
eliminated from our freedom to take 
further residual gauge transformations.  
Thus gravitational degrees of freedom are $D(D+1)/2-1-D-(D-1)=D(D-3)/2$.  
We have no 
gravitational degrees of freedom in D=3.  Without loss of 
generality, the Kalb--Ramond perturbation ${\cal H}_{rtx}$ 
may be taken to be zero 
in the black string background\cite{Gre94}.  
Hence the physical degree of freedom in the 3D  balck string turns out to be 
the dilaton field.  As we will see later, however, the 
conventional (harmonic)gauge is not appropriate for studying the 
propagation of dilaton.

Here we introduce a new gauge (dilaton gauge) to investigate the 
propagation of dilaton.  The harmonic gauge is originated from 
the harmonic coordinates which satisfy
\begin{equation}
g^{\mu\nu} \Gamma^\lambda_{\mu\nu} = 0.
\label{harm_coord}
\end{equation}
Linearization of (\ref{harm_coord}) leads to
\begin{equation}
h^{\mu\nu} \Gamma^\lambda_{\mu\nu}  
      - \bar g^{\mu\nu} \delta \Gamma^\lambda_{\mu\nu}=0
\label{harm_cond}
\end{equation}
with 
$\bar g^{\mu\nu} \delta \Gamma^\lambda_{\mu\nu} = 
\bar \nabla_\mu \hat h^{\mu\lambda}$.  
For the Minkowski background ($\bar g^{\mu\nu} = \eta^{\mu\nu}$), 
(\ref{harm_cond}) becomes the harmonic gauge 
($\bar \nabla_\mu \hat h^{\mu\lambda} =0$)\cite{Wei72}.  But we may choose 
a new gauge 
$h^{\mu\nu} \Gamma^\lambda_{\mu\nu} = \bar \nabla_\mu {\hat h}^{\mu\lambda}$ 
when $\bar g^{\mu\nu} \ne \eta^{\mu\nu}$, 
instead of the harmonic gauge.  In our 
coordinate system, one finds 
\begin{equation}
\bar g^{\mu\nu} \Gamma^t_{\mu\nu}=0,~~
\bar g^{\mu\nu} \Gamma^x_{\mu\nu}=0,~~
\bar g^{\mu\nu} \Gamma^r_{\mu\nu}=-{8 \over {kr}} ( r^2 -MN),
\label{harm_rtx}
\end{equation}
which means that $(t,x,r)$ do not belong to the harmonic coordinates.  
Requiring that (\ref{harm_rtx}) remain unchanged at the linearized level, one 
always choose the new gauge($h^{\mu\nu} \Gamma^\lambda_{\mu\nu} = 
\bar \nabla_\mu {\hat h}^{\mu\lambda}$).  This gauge is suitable for studying 
the propagation of dilaton.

\section{Propagation of string fields}
First we consider the symmetries of the background space--time in 
(\ref{bck_metric}).  
In our case,  the $t$ and $x$--translational symmetries means that 
we can decompose $h_{\mu\nu}$ into frequency modes in these 
variables\cite{Gre94}.  
Hence we take the form for $h_{\mu\nu}$,
\begin{equation}
h_{\mu\nu}(t,x,r) = e^{i \omega t} e^{i \mu x}
\left [
\begin{array}{ccc}
H_{tt}(r)&H_{tx}(r)&H_{tr}(r)\\
H_{xt}(r)&H_{xx}(r)&H_{xr}(r)\\
H_{rt}(r)&H_{rx}(r)&H_{rr}(r)\\
\end{array}
\right ].
\label{ptr_metric}
\end{equation}
Similarly, one chooses the perturbations for Kalb--Ramond 
field and dilaton as
\begin{eqnarray}
{\cal H}_{rtx}(t,x,r) &&= \bar H_{rtx}(r) {\cal H}(t,x,r) 
 =\bar H_{rtx}(r) e^{i \omega t} e^{i \mu x} \tilde{\cal H}(r),
\label{ptr_anti} \\
\phi(t,x,r)&&=e^{i \omega t} e^{i \mu x} \tilde\phi(r).
\label{ptr_scalar}
\end{eqnarray}
Since we start with full degrees of freedom (\ref{ptr_metric}), we 
should choose a gauge to study the propagation of string fields.

\subsection{Harmonic Gauge}
With the harmonic gauge condition(
$\bar\nabla_\mu \hat h^{\mu \rho} = 0$),
the equations (\ref{lin_graviton})--(\ref{lin_anti}) become

\begin{eqnarray}
\bar \nabla^2 h_{\mu \nu} - \bar R_{\sigma \nu} h_{\mu}^{~\sigma}
- \bar R_{\sigma \mu} h_{~\nu}^{\sigma}
+2 \bar R_{\sigma \mu \rho \nu}h^{\sigma \rho}
+2  \bar \nabla_\mu \bar \nabla_\nu \phi  &&\nonumber \\ 
-2 \delta \Gamma^\rho_{\mu\nu} (h) \bar \nabla_\rho \bar \Phi
+ \bar H_{\mu \rho \sigma} {\cal H}_\nu^{~ \rho \sigma}   
- \bar H_{\mu \rho \sigma} \bar H_{\nu\alpha}^{~~\sigma} h^{\rho \alpha} &=& 0,
\label{harmonic_graviton} \\
h^{\mu\nu} \bar \nabla_\mu \bar \nabla_\nu \bar \Phi   
    +h^{\mu \nu}\nabla_\mu \Phi \bar \nabla_\nu \Phi
    - \bar \nabla^2 \phi 
    -2 g^{\mu \nu} \bar \nabla_{\mu}\Phi\bar \nabla_{\nu}\phi
    &&        \nonumber \\
+ {1 \over 6} \Big \{ 2 \bar H_{\mu \rho \sigma} 
                      {\cal H}^{\mu \rho \sigma}                
  - 3 \bar H_{\mu \rho \sigma}   \bar H^{\alpha \rho \sigma} h^\mu_{~\alpha} 
                                    \Big \} &=& 0, 
\label{harmonic_scalar} \\
 (\bar \nabla_\mu + \partial_\mu \bar \Phi) {\cal H}^{\mu \nu \rho}
   - (\bar \nabla_\mu h_\beta^{~\nu}){\bar H}^{\mu \beta \rho}
 + (\bar \nabla_\mu h_\beta^{~\rho}){\bar H}^{\mu \beta \nu}
 + (\partial_\mu \phi) {\bar H}^{\mu \nu \rho}
 &=& 0, 
\label{harmonic_anti}
\end{eqnarray}
Hereafter we are interested only in the dilaton propagation.  
From (\ref{harmonic_scalar}) one finds
\begin{eqnarray}
&&\phi'' + \left( {1 \over r} + {1 \over {r-M}} + {1 \over {r-N}} \right) \phi'
+{kr \over {8(r-M)(r-N)}} \left ( {\omega^2 \over {r-M}} -
{\mu^2 \over {r-N}}\right ) \phi
 \nonumber \\
&&~~~~~~-{1 \over {2r^2(r-M)(r-N)}} \left [ M (r+N) h^t_{~t} + N(r+M) h^x_{~x}
\right . \nonumber \\
&&~~~~~~~~~~~~~~~~~~~~~~~~~~~~~~~~~~~~~\left . 
       + \left \{2r^2-(M+N)r + 2 MN\right \} h^r_{~r} -4 MN {\cal H} \right ]
=0,
\label{harmonic_scalar1}
\end{eqnarray}
where the prime($'$) means the derivative with respect to $r$.
We note that the analysis of dilaton fluctuation around the black 
string reveals a surprising mixing between the dilaton and other fields.  
We have to disentangle this mixing and obtain one decoupled dilaton 
equation by using the harmonic gauge($\bar \nabla_\mu \hat h^{\mu\rho}=0$) 
and Kalb--Ramond equation (\ref{harmonic_anti}).  
However we fail to obtain a decoupled dilaton equation.

\subsection{Dilaton Gauge}
We recognize that it is not easy to decouple the dilaton equation with the 
harmonic gauge condition.  But if we choose the 
dilaton gauge( 
$h^{\mu \nu} \Gamma^\rho_{\mu \nu} = 
\bar\nabla_\mu \hat h^{\mu \rho}$), 
the dilaton equation can be diagonalized easily.
Under this gauge the equations (\ref{lin_graviton})--(\ref{lin_anti}) are 
given by

\begin{eqnarray}
\bar \nabla^2 h_{\mu \nu} - \bar R_{\sigma \nu} h_{\mu}^{~\sigma}
- \bar R_{\sigma \mu} h_{~\nu}^{\sigma}
+2 \bar R_{\sigma \mu \rho \nu}h^{\sigma \rho}
+2  \bar \nabla_\mu \bar \nabla_\nu \phi  
-\bar g_{\mu \sigma} \bar \nabla_\nu ( h^{\alpha \beta} 
             \Gamma^\sigma_{\alpha\beta})
-\bar g_{\nu \sigma} \bar \nabla_\mu ( h^{\alpha \beta} 
            \Gamma^\sigma_{\alpha\beta})
   && \nonumber \\ 
-2 \delta \Gamma^\rho_{\mu\nu} (h) \bar \nabla_\rho \bar \Phi
+ \bar H_{\mu \rho \sigma} {\cal H}_\nu^{~ \rho \sigma}   
- \bar H_{\mu \rho \sigma} \bar H_{\nu\alpha}^{~~\sigma} h^{\rho \alpha} = 0,&&
\label{dilaton_graviton}  \\
h^{\mu\nu} \bar \nabla_\mu \bar \nabla_\nu \bar \Phi   
    +h^{\mu \nu}\bar \nabla_\mu \Phi \bar \nabla_\nu \Phi
  +\bar g^{\mu \nu} \delta\Gamma^\rho_{\mu\nu} \partial_\rho \bar \Phi
    - \nabla^2 \phi 
    -2 \bar g^{\mu \nu} \bar \nabla_{\mu}\Phi\bar \nabla_{\nu}\phi
    ~~~~~~~~~~&&   \nonumber \\
+ {1 \over 6} \Big \{ 2 \bar H_{\mu \rho \sigma} 
                      {\cal H}^{\mu \rho \sigma}                
  - 3 \bar H_{\mu \rho \sigma}   \bar H^{\alpha \rho \sigma} h^\mu_{~\alpha} 
                                    \Big \} = 0,&& 
\label{dilaton_scalar} \\
 (\bar \nabla_\mu + \partial_\mu \bar \Phi) {\cal H}^{\mu \nu \rho}
   - \bar \nabla_\mu h_\beta^{~\nu}{\bar H}^{\mu \beta \rho}
 + \bar \nabla_\mu h_\beta^{~\rho}{\bar H}^{\mu \beta \nu}
 + \partial_\mu \phi {\bar H}^{\mu \nu \rho}
-h^{\delta \eta}\Gamma^\alpha_{\delta\eta} {\bar H}_\alpha^{~\nu\rho}
 = 0.&& 
\label{dilaton_anti}
\end{eqnarray}
Thanks to the dilaton gauge, the first three terms in (\ref{dilaton_scalar}) 
cancel out.  Then the dilaton equation (\ref{dilaton_scalar}) leads to 
\begin{eqnarray}
&&\phi'' + \left( {1 \over r} + {1 \over {r-M}} + {1 \over {r-N}} \right) \phi'
\nonumber \\
&&~~~+{kr \over {8(r-M)(r-N)}} \left ( {\omega^2 \over {r-M}} -
{\mu^2 \over {r-N}}\right ) \phi
-{MN \over {r^2 (r-M)(r-N)}} (h -2 {\cal H}) =0.
\label{eq_scalar}
\end{eqnarray}
Now we attempt to disentangle the final term in (\ref{eq_scalar}) by 
using the dilaton gauge (\ref{harm_cond}) and 
Kalb--Ramond equation (\ref{dilaton_anti}).
Each component of dilaton gauge condition gives rise to 
\begin{eqnarray}
t&:& (\partial_r - {1 \over r} ) h^{tr} + i \omega h^{tt} + i \mu h^{tx} = 0, 
\label{eq_gauge_t}  \\
x&:& (\partial_r - {1 \over r} ) h^{xr} + i \omega h^{xt} + i \mu h^{xx} = 0, 
\label{eq_gauge_x}  \\
r&:& (\partial_r - {1 \over r} ) h^{rr} + i \omega h^{rt} + i \mu h^{rx} = 0. 
\label{eq_gauge_r}
\end{eqnarray}
And the Kalb--Ramond equation (\ref{dilaton_anti}) leads to 
\begin{eqnarray}
tx:&& r (\phi'+{\cal H}' -{h^t_{~t}}' -{h^x_{~x}}')
           + {8 \over k} (r^2-MN)h_{rr} +i\omega h^t_{~r} +i\mu h^x_{~r}=0,
\label{eq_anti_tx} \\
tr:&& i\mu (\phi+{\cal H} -h^t_{~t} -h^r_{~r})
           - {8 \over k} (r-M)h_{xr} +i\omega h^t_{~x} + {h^r_{~x}}' =0,
\label{eq_anti_tr} \\
xr:&& i\omega (\phi+{\cal H} -h^x_{~x} -h^r_{~r})
           - {8 \over k} (r-N)h_{tr} +i\mu h^x_{~t} + {h^r_{~t}}' =0.
\label{eq_anti_xr}
\end{eqnarray}
Solving six equations (\ref{eq_gauge_t})--(\ref{eq_anti_xr}), one finds an 
important equation
\begin{equation}
\partial_\mu (\phi +{\cal H} - h) =0.
\label{eq_anti_simple}
\end{equation}
The solution to this is given by 
\begin{equation}
h=\phi+{\cal H}.
\label{eq_phi}
\end{equation}
This means that the trace of $h_{\mu\nu}(h)$ is a redundant field.  
On the other hand, 
the Einstein equations from (\ref{dilaton_graviton}) are given by
\begin{eqnarray}
tt&:& {8 \over k} (r-M)(r-N) h_{tt}'' 
      + {8 \over k} { {(2r^2-(3M+N)r +2 MN)} \over r} h_{tt}'
      + r \left ( {\omega^2 \over {r-M}} -{\mu^2 \over{r-N}} \right ) h_{tt}
\nonumber \\
&&~~~ +{8 \over k} {{M^2 (r-N)} \over {r^2(r-M)}} h_{tt}
      -{8 \over k} {{MN (r-M)} \over {r^2(r-N)}} h_{xx}
      -{64 \over k} {{M(r-M) (r-N)(r^2-MN)} \over {kr^3}} h_{rr}
\nonumber \\
&&~~~
      -{16 \over k} i\omega {{(r-M) (r-N)} \over {r}} h_{tr}
\nonumber \\
&&~~~
      -{8 \over k} {{M(r-M) (r-N)} \over {r^2}} \phi'
      -2 \omega^2\phi
      + 16 {{MN(r-M)} \over {k r^3}}{\cal H}
      =0,
\label{eq_grav_tt} \\
xx&:& {8 \over k} (r-M)(r-N) h_{xx}'' 
      + {8 \over k} { {(2r^2-(M+3N)r +2 MN)} \over r} h_{xx}'
      + r \left ( {\omega^2 \over {r-M}} -{\mu^2 \over{r-N}} \right ) h_{xx}
\nonumber \\
&&~~~ 
      -{8 \over k} {{MN (r-N)} \over {r^2(r-M)}} h_{tt}
      +{8 \over k} {{N^2 (r-M)} \over {r^2(r-N)}} h_{xx}
      +{64 \over k} {{N(r-M) (r-N)(r^2-MN)} \over {kr^3}} h_{rr}
\nonumber \\
&&~~~
      -{16 \over k} i\mu {{(r-M) (r-N)} \over {r}} h_{xr}
\nonumber \\
&&~~~
      +{8 \over k} {{N(r-M) (r-N)} \over {r^2}} \phi'
      -2 \mu^2\phi
      - 16 {{MN(r-N)} \over {k r^3}}{\cal H}
      =0,
\label{eq_grav_xx} \\
rr&:& {8 \over k} (r-M)(r-N) h_{rr}'' 
      + {16 \over k} { {(3r^2-(M+N)r - MN)} \over r} h_{rr}'
    + r \left ( {\omega^2 \over {r-M}} -{\mu^2 \over{r-N}} \right ) h_{rr}
\nonumber \\
&&~~~
      -{M \over {(r-M)^2}} h_{tt}'
      +{N \over {(r-N)^2}} h_{xx}'
     - {{M(M-N)} \over {(r-M)^3 (r-N)}} h_{tt}
     - {{N(M-N)} \over {(r-M) (r-N)^3}} h_{xx}
\nonumber \\
&&~~~
     +{8 \over k} {{(6r^4-5(M+N)r^3+4MNr^2-MN(M+N)r +2M^2N^2)} \over 
            {r^2(r-M)(r-N)}} h_{rr}
\nonumber \\
&&~~~
      +2 i \omega {M \over {(r-M)^2}} h_{tr}
      -2 i\mu {{N} \over {(r-N)^2}} h_{xr}
\nonumber \\
&&~~~
   +2 \phi''
   +\left( {1 \over {r-M}} + {1 \over {r-N}} \right ) \phi'
      - {{2MN} \over {r^2(r-M)(r-N)}}{\cal H}
      =0,
\label{eq_grav_rr} \\
tx&:& {8 \over k} (r-M)(r-N) h_{tx}'' 
      + {16 \over k} { {(r-M)(r-N)} \over r} h_{tx}'
      + r \left ( {\omega^2 \over {r-M}} -{\mu^2 \over{r-N}} \right ) h_{tx}
\nonumber \\
&&~~~ 
      -{8 \over k} i \mu {{(r^2 -2Nr+MN)} \over r} h_{tr}
      -{8 \over k} i \omega {{(r^2-2Mr+MN)} \over r} h_{xr}
      -2 \omega \mu \phi
      =0,
\label{eq_grav_tx} \\
tr&:& {8 \over k} (r-M)(r-N) h_{tr}'' 
      + {8 \over k} { {(r-M)(3r-N)} \over r} h_{tr}'
      + r \left ( {\omega^2 \over {r-M}} -{\mu^2 \over{r-N}} \right ) h_{tr}
\nonumber \\
&&~~~ 
      +{8 \over k} {{(r-M)} \over r} h_{tr}
      - i \mu {{N} \over {(r-N)^2}} h_{tx}
\nonumber \\
&&~~~ 
      +{{i \omega} \over 2} {{M} \over {(r-M)^2}} h_{tt}
      +{{i \omega} \over 2} {{N} \over {(r-N)^2}} h_{xx}
      +{{4 i \omega} \over k} {{\{(3M+N)r -4 MN\}} \over {r}} h_{rr}
\nonumber \\
&&~~~ 
      +2 i \omega \phi'
      -i \omega {M \over {r(r-M)}}\phi
      =0,
\label{eq_grav_tr} \\
xr&:& {8 \over k} (r-M)(r-N) h_{xr}'' 
      + {8 \over k} { {(3r-M)(r-N)} \over r} h_{xr}'
      + r \left ( {\omega^2 \over {r-M}} -{\mu^2 \over{r-N}} \right ) h_{xr}
\nonumber \\
&&~~~ 
      +{8 \over k} {{(r-N)} \over r} h_{xr}
      + i\omega {M \over{(r-M)^2}} h_{tx}
\nonumber \\
&&~~~ 
      -{{i \mu} \over 2} {{M} \over {(r-M)^2}} h_{tt}
      -{{i \mu} \over 2} {{N} \over {(r-N)^2}} h_{xx}
      +{{4 i \mu} \over k} {{\{(M+3N)r -4 MN\}} \over {r}} h_{rr}
\nonumber \\
&&~~~ 
      +2 i \mu \phi'
      -i \mu {N \over {r(r-N)}}\phi
      =0.
\label{eq_grav_xr} 
\end{eqnarray}
From (\ref{eq_anti_tx})--(\ref{eq_anti_xr}) and 
(\ref{eq_grav_tt})--(\ref{eq_grav_rr}) we may set ${\cal H} = 0$ 
without loss of generality. 
Also this is confirmed from Ref. \cite{Gre94}.
With ${\cal H} = 0$, (\ref{eq_scalar}) becomes a decoupled dilaton equation
\begin{eqnarray}
&&\tilde \phi'' 
+\left ( {1 \over r} + {1 \over {r-M}} + {1 \over {r-N}} \right ) \tilde \phi'
\nonumber \\
&&~~~+{kr \over {8(r-M)(r-N)}} \left ( {\omega^2 \over {r-M}} -
{\mu^2 \over {r-N}}\right ) \tilde \phi
-{MN \over {r^2 (r-M)(r-N)}} \tilde \phi =0.
\label{eq_decoupled}
\end{eqnarray}

\subsection{Stability Analysis}
Since the physical field in the 3D black string is dilaton, 
we will consider the stability of dilaton.  
We remind the reader that the stability analysis should be based on the 
physical fields.  
In order to check whether there exists 
an exponentially growing mode, we replace $\omega$ by $\omega = i \Omega$ in 
(\ref{eq_decoupled})\cite{Reg57}.  
Now we investigate the behavior of $\tilde \phi$ as 
$r \rightarrow \infty$ and as $r \rightarrow M$.  In the asymptotically 
flat region ($r\rightarrow \infty,~k \rightarrow \infty$), one finds the 
relevant equation
\begin{equation}
{{d^2 \tilde \phi} \over {d \rho^2}} - (\Omega^2 + \mu^2) \tilde \phi=0
\label{eq_asymp}
\end{equation}
with the new coordinate 
$\rho = \sqrt{k \over 8} \ln \left ( r \sqrt{2 \over k}\right )$.  
The regular solution to (\ref{eq_asymp}) is given by
\begin{equation}
\tilde \phi_\infty \sim e^{-\sqrt{\Omega^2 + \mu^2} \rho}.
\label{phi_asymp}
\end{equation}
On the other hand, near the horizon ($r \rightarrow M$) we obtain
\begin{equation}
{\tilde \phi_M}'' - 
{{kM \Omega^2} \over {8(M-N)}} { \tilde \phi \over {(r-M)^2}} =0.
\label{eq_horizon}
\end{equation}
Here one finds the regular solution
\begin{equation}
\tilde \phi_M \sim (r-M)^{\sqrt{kM \over {8(M-N)}}\,\Omega}.
\label{phi_horizon}
\end{equation}
It turns out that a necessary condition to obtain a 
regular solution in the whole region is to have a change in sign in 
the coefficient of the undifferentiated $\phi$
in (\ref{eq_decoupled})\cite{Gre94,Reg57}. 
It is obvious that this coefficient does not change sign when one moves 
from $r=M$ to $r=\infty$.  Therefore we conclude that the 3D black 
string is stable.  Also $N=0$ case (neutral black string) cannot 
lead to an instability.

\section{Absorption cross--section}
In this section we will calculate the absorption cross--section 
to obtain the dynamic behavior of 
the black string.  The low energy condition ($\omega \ll {1 \over M}$) 
is assumed, 
which implies that the Compton wavelength of the particle is much larger than 
the gravitational radius of black string.  
Since it is hard to find a 
solution to (\ref{eq_decoupled}), 
we consider the case of $N=0$(neutral black string).  
We also assume 
$\mu \le \omega$  
and use a matching procedure. 
The space--time is divided into two regions: the near region ($r \sim M$) 
and far region ($r \gg {1 \over \omega}$)\cite{Str97}.  
We now study each region in turn.

\subsection{Near--Region Solution}
In this case, (\ref{eq_decoupled}) leads to 
\begin{equation}
\tilde \phi '' + \left ( { 2\over r} + {1 \over {(r-M)}} 
\right ) \tilde \phi ' + {1 \over 8} {k \over {r-M}}
\left ( {\omega^2 \over {r-M}} - {\mu^2 \over {r}} \right ) \tilde \phi=0.
\label{eq_near}
\end{equation}
In order to solve the above equation, we introduce a new variable 
\begin{equation}
z = 1 - {M \over r},~~0 \le z \le 1.
\label{def_z}
\end{equation}
The horizon is located at $z=0$ and the asymptotically flat region is 
at $z=1$.  Then (\ref{eq_near}) is given by
\begin{equation}
z(1-z) \partial_z^2 \tilde \phi  
+ \partial_z \tilde \phi  
+ {k \over 8} \left ( {\omega^2 \over z} + 
             {{\omega^2-\mu^2} \over {1-z}} \right ) \tilde \phi=0.
\label{eq_near_z}
\end{equation}
This can be transformed into the hypergeometric form by defining
\begin{equation}
\tilde \phi = z^\alpha (1-z)^\beta \Psi.
\label{def_psi}
\end{equation}
(\ref{eq_near_z}) leads to
\begin{equation}
z(1-z) \partial_z^2 \Psi  
+ \{ 1 + 2\alpha - 2(\alpha +\beta)z \} \partial_z \Psi  
- ( \alpha+\beta)( \alpha+\beta-1) \Psi=0
\label{eq_near_psi}
\end{equation}
where
\begin{equation}
\alpha = i \sqrt{k \over 8} \omega, 
~ \beta = 1 + i \sqrt{{k\over 8}(\omega^2 -\mu^2) -1} .
\label{alpha_beta}
\end{equation}
Comparing (\ref{eq_near_psi}) with the standard hypergeometric 
equation, one finds the solution
\begin{eqnarray}
\tilde \phi(z) =&& z^\alpha(1-z)^\beta \left [
 C_1 F(\alpha+\beta,\alpha+\beta-1, 1+ 2 \alpha ; z)  \right .
\nonumber \\
 &&~~~~~~~~\left . 
 +C_2 z^{-2 \alpha} F(-\alpha+\beta,-\alpha+\beta-1, 1- 2 \alpha ; z) 
\right ]
\label{sol_near}
\end{eqnarray}
where $C_1$ and $C_2$ are to--be--determined constants.
Near the horizon ($z \rightarrow 0, ~r\rightarrow M$), one gets
\begin{equation}
\tilde \phi_M(z) = 
{C_1 \over M^\alpha} e^{i \sqrt{ k \over 8} \omega \ln(r-M)}
+{C_2 M^\alpha} e^{-i \sqrt{ k \over 8} \omega \ln(r-M)},
\label{sol_horizon}
\end{equation}
where the first term (last term) correspond to the ingoing(outgoing) 
wave.  We impose the condition that there be only ingoing flux at the 
horizon to obtain the absorption cross--section.  This implies 
$C_2=0$ and thus the solution around $z=0$ is
\begin{equation}
\tilde \phi_{\rm near}(z) = C_1 z^\alpha(1-z)^\beta 
 F(\alpha+\beta,\alpha+\beta-1, 1+ 2 \alpha ; z) . 
\label{sol_near0}
\end{equation}

\subsection{Far--Region Solution}
In this region (\ref{eq_near}) reduces to
\begin{equation}
\tilde \phi '' + { 3\over r}  
 \tilde \phi ' + {k \over 8} 
{{\omega^2-\mu^2} \over r^2} \tilde \phi=0.
\label{eq_far}
\end{equation}
The solution of this equation is 
\begin{eqnarray}
\tilde \phi_{\rm far}(r) = &&
  A \left ( { M \over r} \right ) ^{2 -\beta} 
+ B \left( { M \over r} \right ) ^{\beta} \nonumber \\
=&& {1 \over r} \left [
A M^{2-\beta}e^{i \sqrt{{k \over 8} (\omega^2-\mu^2) -1}\, \ln r}
+B M^\beta e^{-i \sqrt{{k \over 8} (\omega^2-\mu^2) -1} \, \ln r} \right ],
\label{sol_far}
\end{eqnarray}
where $A$ and $B$ are the normalization constants.  Here the first 
(last) term correspond to the ingoing (outgoing) wave in the 
asymptotically flat region.  

\subsection{Matching the far and near solutions}
Now we need to match the far--region solution (\ref{sol_far}) to the 
large $r$($z \rightarrow 1$) limit of near--region solution 
(\ref{sol_near0}) in the overlapping region. The $z\rightarrow 1$ 
behavior of (\ref{sol_near0}) follows from the $z\rightarrow 1-z$ 
transformation rule for hypergeometric functions. This takes the form
\begin{eqnarray}
\tilde \phi_{n \rightarrow f}(z) &&= 
C_1 z^\alpha (1-z)^\beta \left [
{{\Gamma(1+2 \alpha) \Gamma(2 -2\beta)} \over 
    {\Gamma(1+ \alpha -\beta ) \Gamma(2 +\alpha -\beta)}}
    F(\alpha+\beta,\alpha+\beta-1,2 \beta -1; 1-z)
\right . \nonumber \\
&&\left . 
+ (1-z)^{2-2\beta} {{\Gamma(1+2 \alpha) \Gamma(-2 +2\beta)} \over 
    {\Gamma(\alpha+\beta) \Gamma(\alpha +\beta-1)}}
    F(\alpha-\beta+1,\alpha-\beta+2,3-2\beta ; 1-z)
\right ].
\label{near2far}
\end{eqnarray}
Using $1-z = {M \over r}$, one obtains the explicit form
\begin{eqnarray}
\tilde \phi_{n\rightarrow f}(r) = &&
C_1 \left ({M \over r} \right )^\beta 
{{\Gamma(1+2 \alpha) \Gamma(2 -2\beta)} \over 
    {\Gamma(1+ \alpha -\beta) \Gamma(2 +\alpha -\beta)}}
+C_1 \left ( {M \over r} \right )^{2-\beta}
{{\Gamma(1+2 \alpha) \Gamma(-2 +2\beta)} \over 
    {\Gamma(\alpha+\beta) \Gamma(\alpha +\beta-1)}}.
\label{sol_near2far}
\end{eqnarray}
Matching (\ref{sol_far}) to (\ref{sol_near2far}), we find
\begin{eqnarray}
A &=& 
{{\Gamma(1+2 \alpha) \Gamma(-2 +2\beta)} \over 
    {\Gamma(\alpha+\beta) \Gamma(\alpha +\beta-1)}}C_1,
\nonumber \\
B &=& 
{{\Gamma(1+2 \alpha) \Gamma(2 -2\beta)} \over 
    {\Gamma(1+ \alpha-\beta ) \Gamma(2 +\alpha -\beta)}}C_1 .
\label{sol_AB}
\end{eqnarray}
The reflection coefficient ${\cal R}$ is then given by
\begin{equation}
{\cal R} = \left | { B \over A} \right | ^2 =
\left |
{{\Gamma(\alpha+\beta) \Gamma(\alpha+\beta-1)} \over 
    {\Gamma(1+ \alpha-\beta) \Gamma(2 +\alpha -\beta)}}
\right |^2.
\label{coef_ref}
\end{equation}
The absorption coefficient ${\cal A}$(partial wave absorption 
cross--section) is calculated as
\begin{eqnarray}
{\cal A} =&& 1 -{\cal R}
\nonumber \\
=&&
{{\sinh{2 \pi \sqrt{k \over 8} \omega} 
  \sinh{2 \pi \sqrt{{k \over 8}(\omega^2-\mu^2)-1} }}
\over 
{\sinh^2{ \pi \left (\sqrt{k \over 8} \omega+
      \sqrt{{k \over 8}(\omega^2-\mu^2)-1}\right )}}}
.
\label{coef_abs}
\end{eqnarray}
This result can also be obtained from the flux calculation.  
The conserved flux due to (\ref{eq_near})
is defined as 
\begin{equation}
{\cal F} = { 2\pi \over i} 
\left ( 
\tilde \phi^* \Delta \partial_r \tilde \phi
-\tilde \phi \Delta \partial_r \tilde \phi^*
\right ).
\label{flux_horizon}
\end{equation}
Using (\ref{sol_near0}) and $\Delta = r^2(r-M)$, the 
incoming flux across the horizon is given by
\begin{equation}
{\cal F}(0) = {4\pi M^2} \sqrt{k \over 8} |C_1|^2 \omega.
\label{fluz_horizon0}
\end{equation}
From (\ref{sol_far}) the incoming flux at infinity is 
found to be
\begin{equation}
{\cal F}_{in} =  4\pi |A|^2 M^2  \sqrt{{k \over 8}(\omega^2 -\mu^2) -1}.
\label{flux_infinity}
\end{equation}
Hence the partial wave absorption cross--section is
\begin{eqnarray}
\sigma_{\rm BS}^\phi = && {{{\cal F}(0)} \over {{\cal F}_{in}}}
\nonumber \\
=&& \sqrt{k \over 8}
{ \omega \over \sqrt{{k \over 8}(\omega^2 -\mu^2)-1}}
\left |
{{\Gamma(\alpha+\beta) \Gamma(\alpha+\beta-1)} \over 
    {\Gamma(1+ 2\alpha) \Gamma(2 \beta-2)}}
\right |^2
\nonumber \\
=&&
{{\sinh{2 \pi \sqrt{k \over 8} \omega} 
  \sinh{2 \pi \sqrt{{k \over 8}(\omega^2-\mu^2)-1} }}
\over 
{\sinh^2{ \pi \left (\sqrt{k \over 8} \omega+
      \sqrt{{k \over 8}(\omega^2-\mu^2)-1}\right )}}}
\label{x_section}
\end{eqnarray}
We note that in the case of 3D black string the 
absorption coefficient(${\cal A}=\sigma_{\rm BS}^\phi $) 
is just the plane--wave absorption cross--section. 

\section{Discussions}
First, let us discuss the importance of our new gauge(dilaton gauge).  
It turns out that the conventional(harmonic) gauge is not appropriate 
for investigating the propagation of dilaton.  Instead, here we 
introduce the dilaton gauge.  This gauge transforms the graviton 
part into complicated form, while it simplifies the mixing between 
the dilaton and the other fields.  Further, the dilaton gauge condition 
itself provides us with the simple relations 
(\ref{eq_gauge_t})--(\ref{eq_gauge_r}).  This is so because 
$h^{\mu\nu}\Gamma^\rho_{\mu\nu} = \bar \nabla_\mu \hat h^{\mu\rho}$ is 
reduced to $\partial_\mu h^{\mu\rho}= {1 \over r} h^{r\rho} $.  
And the Kalb--Ramond equation is also simplified as in 
(\ref{eq_anti_tx})--(\ref{eq_anti_xr}) by this gauge.  We note that the 
trace of $h_{\mu\nu}(h)$ is not zero but a redundant one ($h=\phi+{\cal H}$) 
under this gauge.  Thanks to the dilaton gauge, we obtain the 
decoupled dilaton equation.  The stability analysis based on this equation 
shows that the 3D black string is stable.

Now we turn our attention to the dynamic behavior of neutral black string.  The 
relevant quantity is the absorption cross--section.  In the case of charged 
black string($N \ne 0$), we cannot obtain this quantity.  For the low 
energy ($\omega \ll {1 \over M}$) and $\omega \ge \mu$, one finds the 
absorption cross--section using a matching procedure.  Since the dilaton 
belongs to the fixed scalar\cite{Cal97}, it is useful to compare the absorption 
cross--section (\ref{x_section}) with that of 
minimally coupled scalar($\psi$)\cite{Mal97}. This field satisfies the 
equation: $\bar \nabla^2 \psi=0$. By similar procedure, one can obtain its 
cross--section as
\begin{equation}
\sigma_{\rm BS}^\psi = 
{{\sinh{2 \pi \sqrt{k \over 8} \omega} 
  \sinh{2 \pi \sqrt{{k \over 8}(\omega^2-\mu^2)} }}
\over 
{\sinh^2{ \pi \left (\sqrt{k \over 8} \omega+
      \sqrt{{k \over 8}(\omega^2-\mu^2)}\right )}}}.
\label{x_section_min}
\end{equation}
If ${k \over 8} (\omega^2-\mu^2) \gg 1$, there is essentially 
no difference between 
(\ref{x_section}) and (\ref{x_section_min}).  This means that to 
extract the dynamic behavior of the neutral($N=0$) black string, one can use 
either the minimally coupled scalar or the dilaton (fixed scalar). 
Further let us compare (\ref{x_section_min}) with the absorption cross--section 
of minimally coupled scalar in the BTZ balck hole\cite{Bir97}. 
In the BTZ black hole, 
one finds $\sigma_{\rm BTZ}^\psi \rightarrow A_H(=2 \pi r_+)$ when 
$m=0,~\omega \to 0$. In our case, we find 
the similar result ($\sigma_{\rm BS}^\phi \to 1$: total absorption) 
for $\mu=0,~ \omega \to 0$. 

Finally we comment on the 3D extremal ($M=N$) black string. This was 
discussed in Ref. \cite{Lee97}. In this case the metric is not only 
translationally invariant, but boost invariant.  In other words, the 
space--time of 3D extremal black string has a null Killing symmetry. 
In this case, it turns out that the graviton become a propagating mode, 
whereas the dilaton is non--propagating.  This case contrast well with 
the present case.  However, the graviton may become a propagating mode 
by the transmutation of the degrees of freedom with other field 
(here, dilaton) in the extremal balck string space--time.  This is 
very similar to the Higgs mechanism for gauge fields in the Minkowski 
space--time.

\section*{Acknowledgement}
This work was supported in part by the Basic Science Research Institute 
Program, Minstry of Education, Project NOs. BSRI--97--2441 and 
BSRI--97--2413.

\end{document}